\documentclass{article}

\usepackage[english]{babel}
\usepackage{natbib}

\usepackage[letterpaper,top=2cm,bottom=2cm,left=3cm,right=3cm,marginparwidth=1.75cm]{geometry}

\usepackage{amsmath}
\usepackage{graphicx}
\usepackage[colorlinks=true, allcolors=blue]{hyperref}

\usepackage{xcolor}

\usepackage{blindtext}
\usepackage{geometry}
\newcommand{\gsim}{\mathrel{\rlap{\lower 3pt \hbox{$\sim$}} \raise 2.0pt \hbox{$>$}}}

\newcommand{\Msun}{M$_\odot$ }

\begin{document}



\begin{center}

{\large ESO Expanding Horizons White Paper}

\vspace{2mm}

{\LARGE{\bf Electromagnetic counterparts of massive BH mergers with LISA}}\\

\end{center}

\vspace{3mm}

\noindent{\large \bf Contact Authors:}

\normalsize
{\bf Massimo Dotti}, Universit\'a di Milano Bicocca, Italy, massimo.dotti@unimib.it

{\bf Filippo Mannucci}, INAF-Arcetri, Italy, filippo.mannucci@inaf.it\\

\noindent {\large \bf Proposing team:}

\noindent
{\bf - R. Buscicchio} Univ. Milano Bicocca, Italy
{\bf - M. Colpi} Univ. Milano Bicocca, Italy
{\bf - Q. D'Amato} INAF-Arcetri, Italy,
{\bf - A. De Rosa} INAF-IAPS, Rome, Italy
{\bf - A. Franchini} Univ. Milano, Italy
{\bf - Z. Haiman} Inst. of Science and Technology Austria (ISTA), Klosterneuburg, Austria
{\bf - D. Izquierdo-Villalba} Inst. of Space Sciences (ICE, CSIC), Spain
{\bf - A. Mangiagli} Max Planck Inst. for Gravitational Physics (Albert Einstein Inst.), Germany
{\bf - M. Scialpi} Univ. of Florence, Italy
{\bf - P. Severgnini} INAF-OABrera, Italy
{\bf - C. Vignali} Unive. of Bologna, Italy
{\bf - M. Volonteri} CNRS/Inst. d'Astrophysique de Paris, France
\\

\noindent {\large \bf Supporting team:}

\small
\noindent
{\bf Antonini F.} Cardiff univ., UK
{\bf - Ashall C.}, Inst. for Astronomy, Univ. of Hawai'i, USA
{\bf - Barausse E.}, SISSA, Trieste, IT
{\bf - Beckmann R. S.} Inst. for Astronomy, Univ. of Edinburgh, UK
{\bf - Berti E.}, Johns Hopkins Univ., Baltimore, USA
{\bf - Bianchi S.}, Dip. di Matematica e Fisica, Univ. Roma Tre, Rome, IT
{\bf - Bonetti M.}, Univ. Milano-Bicocca, IT
{\bf -Bortolas E.}, INAF--Osservatorio Astronomico di Padova, IT
{\bf - Bourne M. A.}, Univ. of Hertfordshire, UK
{\bf - Capelo~P.~R.}, Dep. of Astrophysics, Univ. of Zurich, CH
{\bf - Caramete~A.}, Inst. of Space Science – INFLPR Subsidiary , Romania
{\bf - Caramete~L. I.}, Inst. of Space Science – INFLPR Subsidiary , Romania
{\bf - Carleo~A.}, INAF-Osservatorio Astronomico di Cagliari, IT
{\bf - Cook D.}, IPAC, California Inst. of Technology, USA
{\bf - D'Ammando F.}, INAF-IRA Bologna, IT
{\bf - Dainotti M.}, National Astronomical Obs. of Japan, Tokyo, Japan
{\bf - Dayal P.}, Canadian Inst. for Theoretical Astrophysics,  Univ. of Toronto, Canada
{\bf - Destounis K.}, Center for Astrophysics and Gravitation – CENTRA, Dep. de Física, Inst. Superior Técnico – IST, Univ. de Lisboa – P
{\bf - Fritz A.}, Kuffner Obs., Johann-Staud-Straße 10, 1160 Vienna, Austria
{\bf - Gourgouliatos K.N.}, Lab. of Universe Sciences, Dep. of Physics, Univ. of Patras, Greece 
{\bf - Gualandris A.}, Univ. of Surrey, UK
{\bf - Gurvits~L.I.}, Delft Univ. of Technology, NL 
{\bf - Ili\'c~D.I.}, Univ. of Belgrade, Faculty of Mathematics, Serbia
{\bf - Ivanov~V.~D.}, ESO, D
{\bf - Izzo~L.}, INAF, Osservatorio Astronomico di Capodimonte, Naples, IT
{\bf - Jaisawal~G. K.}, DTU Space, Technical Univ. of Denmark, DK
{\bf - Johansson~P. H.}, Dep. of Physics, Univ. of Helsinki, Finland
{\bf - Korol V.}, Max Planck Inst. for Astrophysics, D
{\bf - Kotak R.}, Dept. of Physics \& Astronomy, Univ. of Turku, Finland
{\bf - Kova{\v c}evi\'c~A.B.}, Univ. of Belgrade, Faculty of Mathematics, Serbia
{\bf - Kuncarayakti H.}, Univ. of Turku, Finland
{\bf - Li K.}, the Flatiron Inst.'s Center for Computational Astrophysics, the Simons Foundation, U.S.
{\bf - Liu X.}, Univ. of Illinois at Urbana-Champaign, USA
{\bf - Lupi A.}, Como Lake Center for Astrophysics, Univ. Insubria, IT
{\bf - Mainieri V.}, ESO, D
{\bf - MacBride S.}, Physik Institut, Univ. of Zurich, CH
{\bf - Majidi F. Z.}, INAF OACN, Italy
{\bf - Marconcini C.}, Dip. di Fisica e Astronomia, Univ. di Firenze, IT
{\bf - Mayer L.}, Dep. of Astrophysics, Univ. of Zurich, CH
{\bf - Mignon-Risse R.}, Aix Marseille Univ., CNRS, CNES, LAM, Marseille, F
{\bf - Panda, S.}, Intern. Gemini Obs./NSF NOIRLab, La Serena, Chile
{\bf - Pollo, A.}, National Centre for Nuclear Research, Otwock, Poland 
{\bf - Porquet, D.}, Aix Marseille Univ, CNRS, CNES, LAM, Marseille, F
{\bf - Pozo Nu\~nez, F.}, Astroinformatics, Heidelberg Inst. for Theoretical Studies, Heidelberg, D
{\bf - Ramond, P.}, Lab. de Physique Corpusculaire, ENSICAEN, CNRS/IN2P3, Caen, F
{\bf - Regan J.}, Dep. of Physics, Maynooth Univ., Ireland
{\bf - Roth M.M.}, Deutsches Zentrum f\"ur Astrophysik (DZA), Görlitz, D
{\bf - Runnoe J.~C.}, Vanderbilt Univ., Fisk Univ., USA
{\bf - Ruiz M.}, Dep. Astron. i Astrof. Univ. de València, ES
{\bf - Schneider R.}, Dip. di Fisica, Rome Univ. Sapienza, IT
{\bf - Schussler F.} IRFU, CEA, Univ. Paris-Saclay, F
{\bf - Sesana, A.} Univ. Milano Bicocca, Milan, IT
{\bf - Shaifullah, G.} Univ.i Milano Bicocca, Milan, IT
{\bf - Sopuerta, C.F.}, Inst. of Space Sciences (ICE-CSIC and IEEC), Bellaterra, Barcelona, ES
{\bf - Steinle N.}, Univ. of Manitoba and Winnipeg Inst. for Theoretical Physics, Canada 
{\bf - Svoboda J.}, Astronomical Inst. of the Czech Academy of Sciences, Czechia
{\bf - Szab\'o R.}, HUN-REN CSFK, Konkoly Obs., Hungary
{\bf - Tamanini N.}, Univ. de Toulouse, CNRS/IN2P3, L2IT, Toulouse, F
{\bf - Tintareanu-Mircea O.}, Inst. of Space Science - INFLPR Subsidiary, Romania
{\bf - Tozzi G.}, Max Planck Inst. for Extraterrestrial Physics, D
{\bf - Trinca A.}, Inst. for Astronomy, Univ. of Edinburgh, UK
{\bf - Trotta, R.}, SISSA, Trieste, IT \& Imperial College London, UK
{\bf - Ugolini, C.}, Gran Sasso Science Inst., L'Aquila, IT
{\bf - Urbanec M.}, Inst. of Physics, Silesian University in Opava, Czech Republic
{\bf - Vecchio~A.}, Inst. for GW Astronomy \& School of Physics and Astronomy, Univ. of Birmingham, UK
{\bf - Vernieri~D.}, Univ. Napoli Federico II, Naples, IT
{\bf - Warburton~N.} Univ. College Dublin, Ireland
{\bf - Yazadjiev~S.~S.}, Dep. of Theoretical Physics, Sofia Univ., Bulgaria
{\bf - Zanchettin~M.V.} INAF-Arcetri, IT
{\bf - Zumalacarregui~M.}, Max Planck Inst. for Gravitational Physics (Albert Einstein Inst.) Potsdam, D
{\bf - Tombesi F.}, Rome Tor Vergata Univ., IT

\newpage
\normalsize
\section{Context and scientific questions}

The Laser Interferometer Space Antenna (LISA, Colpi et al., 2024), adopted by ESA
and scheduled for the second half of the next decade, will drive a new revolution
in the rapidly growing field of gravitational-wave (GW) astronomy, by extending GW
observations into the hiterto  unexplored millihertz regime. One of the key source
classes of LISA is  merging  massive black hole binaries (MBHBs) in the $10^4$-$10^7$
\Msun  mass range detectable to very high redshifts $z\sim 15$. 
MBHBs lighter than $10^6$ \Msun can potentially be identified during their
inspiral weeks before coalescence, allowing for complementary electromagnetic
(EM) observations before, during, and after the two MBHs merge.

Joint GW-EM detections would enable for the first time to address a number of crucial science questions. These multimessenger observations have the potential to: 
uncover the behavior  of magnetized plasmas in dynamical strong-field spacetimes; 
set exquisite constraints on the accretion rate and efficiency on MBHs, providing multiple independent contraints on MBH spins; 
allow for an independent measurement of the speed of GWs, constraining the mass of the graviton and possible violations of Lorentz invariance; provide tight constraints on cosmological parameters;
study the environment of the MBHB;
understand the role of MBH mergers in their growth; 
inform our understanding of the coevolution of MBHs and their host galaxies 
(e.g., Bogdanov\'ic et al. 2022 and refs. therein). 
Key to achieving these breakthrough results will be the ability to localize MBHBs
and to define the optimal window for EM observations.

\section{Theoretical predictions of EM signatures}

This white paper aims at defining the optimal strategy to maximize the number of detected EM counterparts of LISA MBHB events.

The prospects for joint EM–GW identification of LISA MBHBs are primarily limited by LISA ability to constrain both the sky-localization region and the luminosity distance using GW observations alone. For a representative system with a total mass of $3\times10^5$ \Msun at $z=1$, the sky-localization accuracy ranges from $\gsim 100$ square degrees when the signal is first detected weeks before merger, drastically improving to 0.01 $\rm deg^2$ at the time of coalescence (e.g. Mangiagli et al. 2020). 
These error boxes typically contain $\sim 10^4$ ($10^3$) galaxies within ten hours (one hour) before coalescence (Lops et al. 2023). 
Unfortunately, the host galaxies  of typical LISA binaries are not expected to exhibit distinctive intrinsic properties that would help in their identification within the broader galaxy population 
(e.g. Volonteri et al.  2020; Izquierdo et al. 2023). 
For this reason, the community has focused primarily on identifying the EM signatures of accreting, coalescing binaries that would make counterpart detection possible. Indeed, during a galaxy merger, tidal and hydrodynamical processes drive substantial gas inflow toward the centre of the two galaxies (e.g., Capelo \& Dotti 2017, and refs. therein). These processes continue to funnel  gas to progressively smaller radii until the two MBHs reach the centre of the remnant galaxy. This scenario suggests that a non-negligible fraction of MBHBs may be accreting at high rates during their inspiral and merger.

If some amount of gas is present in the proximities ($\sim$10 gravitational radii) of the binary close to its coalescence, the time dependent metric evolution can deposit enough energy to result in an Eddington level outburst of energy (Krolik, 2010). The duration and spectrum of such a bursts are, however, very uncertain, depending on the amount of material present in the binary vicinity and on the micro-physical processes occurring within such gas distribution. 
Recent numerical simulations, ranging from 2D pure hydrodynamical to full general relativity radiation-magnetohydrodynamical, have shown that as long as there is sufficient circumbinary gas present,
the binary can accrete nearly as efficiently as a single MBH, regardless of the cavity carved by the binary,  
until $\sim 1$ day before merger (e.g. Franchini et al. 2024, Krauth et al. 2025, Ennoggi et al. 2025). The motion of the MBHs and the time dependent gravitational potential imprint periodicities commensurate to the binary orbital period in the light-curves potentially observable from optical to X-rays (see D'Orazio and Charisi 2023 and references therein). %
At $\sim 1 $ day to the coalescence,
the tidal
field between the two MBHs becomes too large, and the above mentioned simulations find that  disc-mediated accretion onto individual MBHs ceases, resulting in  
 a  drop in the thermal UV/X-ray emission, whose exact magnitude depends on the gas thermodynamics (e.g. Franchini et al. 2024), with a possible increasing contribution from shocks from gas on radial orbits close to the MBHs (e.g. Enoggi et al. 2025).

All of the features mentioned above can be jeopardized by the large error box of LISA detections, which will require rapid and deep observations over large sky areas, as well as by merger-driven gas inflows that, while boosting the probability of AGN activity around a MBHB, also  enhance the probability of obscuration. Designing tailored observational strategies is therefore of paramount importance to guarantee the feasibility of multimessenger studies.

\section{Observing strategy, technology development and data handling requirements 
}

EM signatures of the inspiraling and merging  MBHB can help pinpoint which of the tens of thousands of candidate galaxies host the coalescing binary. \\

{\bf Wide field imagers}: optical/near-IR, instruments like {\bf 
Rubin/LSSTCam} 
(Ivezi{\'c}  et al. 2019) can be used to monitor the patches of sky corresponding to the LISA events with good sky localization 
for days to weeks before merging. 
Such observations can potentially identify periodic sources if the binary is accreting and unobscured. 
Having access to one or more such instruments is crucial for this science case. Since the expect periods are a few minutes or less, short integrations must be used, limiting the sensitivity. Instruments more sensitive than LSST would be ideal to extend detections towards lower mass MBHB. 
Even if any AGN activity shuts off before LISA detects the system, periodic light curves can still be searched for within the LISA error box in the catalogues assembled by large ground-based surveys prior to LISA's launch, such as LSST in the optical, and {\bf ULTRASAT} and its successors in the UV (e.g. Xin et al. 2025 and references therein). These catalogs will enable a selection of promising candidates to follow-up with other targeted  EM observations.
X-ray follow-ups can
be planned on previously identify targets or when the sky localization improves to check for the current level of activity (in X-rays, e.g. by triggering {\bf NewAthena}, {\bf AXIS}, or {\bf THESEUS}
observation (Cruise et al., 2025; Reynolds et al. 2023; Amati et al., 2021). 
Significant dimming of the X-ray flux with respect to historical observations of the field can also be used to pin-point the progenitor system.\\

{\bf Multi-object spectroscopy}: another possible observational strategy could include a fast optical-near infrared spectral analysis of the galaxies within the LISA error box, to search for signs of AGN activity traced by narrow emission lines (see, e.g. Mangiagli et al. 2020). Such effort would require a large telescope (to minimize the exposure time), large field of view (to minimize the number of pointings), and, for a fiber-fed  spectrograph, tens of thousands of fibers, such as the proposed {\bf WST} (Mainieri et al., 2024). The identification of Type II AGN spectra would imply that these
systems were accreting hundreds of years ago (corresponding to the light travel time
from the nucleus to the narrow line region). Targeted EM follow-ups can
be then used to check for the current level of activity (e.g. in X-rays or in radio, as for the candidates selected by wide field imagers), testing the prediction of a
dimming accretion luminosity precursor.
The identification of Type II AGN at the low metallicities expected for the low-mass galaxies hosting most of the LISA events can be 
challenging, and requires the 
analysis of faint UV and optical lines (e.g. Feltre et al., 2026; Mazzolari et al. 2024 and references therein).
The observation of these lines are within the spectroscopic capabilities of optical multi-object spectrographs at 8-12m class telescopes up to redshifts $z\sim 1$ using optical lines and $z\sim 4$ for UV lines. Multi-objects spectrographs covering the near-IR wavelength range, such as {\bf VLT/MOONS} (Cirasuolo et al., 2020) or the proposed near-IR arm of WST, would allow to extend the detection of optical lines up to $z\sim 2$ and greatly extend the cosmic volume covered.
The identification of rarer unobscured binaries can be identified by the presence of
broad emission lines, triggering the search for the variability MBHB features discussed above as well
as for the search of peculiarities in the broad line profiles
(e.g. Bertassi et al. 2025 and references therein). \\

{\bf Radio observations}: the rapid changes in the MBHs intrinsic properties (mass and spin) and of their local environment (magnetic field and gas properties) at coalescence can result in variations in the jet power, hence in either increased or decreased radio luminosity.
Such change can be detected with 
future facilities (e.g. {\bf SKA}), even under very conservative assumptions (see e.g. Dong-P{\'a}ez et al. 2023 and references therein). 
As radio emission is not affected by obscuration, this diagnostic can serve as a valuable tracer of the most dust-enshrouded regions. The SKA combination of high sensitivity, large field of view, and excellent spatial resolution provides unique capabilities for detecting and localizing such events. 
A survey similar to the one proposed in Prandoni \& Seymour (2015, of $\sim 31000 \deg^2$, down to 5 $\mu$Jy/beam sensitivity) can  provide an immediate flux reference (at least for the brightest LISA binaries in radio) for comparison with immediately pre-merger and post-merger observations of the LISA error box, allowing to pin-point candidates even in case the jet shuts-off before LISA detection.\\

Only the detection of {\bf periodic variability} with the right frequency uniquely identifies close to coalescence MBHBs. For example, a drop in the observed X-ray flux might be associated with a change in the hydrogen column density, or might be caused by intrinsic long term variability of single MBH accretion. This second scenario could also explain the dimming of radio jets. An accurate statistical framework to evaluate the probability of having observed a binary must be developed in the next years, taking into account the chances of observing possible EM tracers jointly with the information that (at least) one galaxy in the error box is hosting a coalescing binary (see Dal Canton et al. 2019 for an early example). This strategy will leverage on large AGN surveys available by the time LISA will fly, to characterize the false alarm probability of EM binary features.\\

{\bf Deep follow-ups}: the identification of  one or a small number of host candidates will allow for deeper follow-ups with smaller field of view instruments, as, among others, the {\bf ELT (MICADO, METIS, HARMONI, and MOSAIC,} 
Davies et al 2021; Brandl et al., 2021; Thatte et al., 2021, Hammer et al., 2021), 
the already cited NewAthena, and  {\bf ngVLA} (Murphy et al., 2018).
At the same time, limiting the number of candidates would simplify the search for post-merger signatures, like, for example, the accretion disc (and X-ray corona) re-brightening after the binary coalescence
(Franchini et al. 2024).
The uncertainties on the expected timescales for the occurrence of such processes  are extremely large 
(Krolik et al., 2010, Bogdanov\'ic et al 2022)
and the pre-selection of a few candidate hosts (the LISA error box shrinks significantly when the coalescence is included in the analysis) could allow for long term monitoring campaigns with logarithmically spaced observations in time,  enabling the secure identification of the newly formed MBH remnant and maximizing the scientific payoff of the GW detection.

\section{Between now and the LISA events}

Significant progress has been made over the past decades both in the modeling of MBHB EM signatures and in their identification within real observational data. The characterization of the EM phenomenology of MBHBs is expected to continue improving throughout the next decade. From the observational side, we may obtain securely identified MBHBs even before LISA becomes operational, potentially enabling the detection of binaries coalescing during the mission itself (Xin \& Haiman 2021; Haiman et al. 2023). Also, observations of the progenitors of the merging MBHB both as binary BH (e.g., Karuse et al. 2025) or as dual AGN at kpc separations (e.g., De Rosa et al., 2019; Mannucci et al., 2023, Perna et al., 2025) can contribute in better estimate the number of mergers, currently very uncertain, tailoring more efficient observing strategies.  

With this white paper, we aim to stimulate a timely and urgent discussion on the observational strategies and instrumentation required to achieve the identification of LISA EM counterparts.
\vskip 1 Truecm
{\bf \large \noindent References}\\
\small
\noindent
$\bullet$ Amati, L., et al., 2021, ExA 52, 183;
$\bullet$ Bertassi, L. et al. 2025, A\&A, 702, 165;
$\bullet$ Bogdanov\'ic, T., Coleman Miller, M., \&, and Blecha, L., 2022, Living Rev. in Relat., 25(1):3;
$\bullet$ Brandl, B., et al., 2021, Msngr 182, 22;
$\bullet$ Capelo, P., \& Dotti, M., 2017, MNRAS 465, 2643; 
$\bullet$ Cirasuolo, M., et al., 2020 arXiv:2009.00628;
$\bullet$ Colpi, M., et al., 2024, {\it LISA redbook}, arXiv:2402.07571; 
$\bullet$ Cruise, M., et al., 2025, NatAs, 9, 36;
$\bullet$ Dal Canton et al. 2019, ApJ, 886, 146;
$\bullet$ Davies, R., et al., 2021, Msngr 182, 17;
$\bullet$ De Rosa, A., et al., 2020, NewAR, 8601525D;
$\bullet$ Dong-P{\'a}ez C.~A., et al., 2023, A\&A, 676, A2;
$\bullet$ D'Orazio D.~J., Charisi M., 2023, arXiv:2310.16896;
$\bullet$ Ennoggi, L., et al., 2025, PhRvD, 112, 063009;
$\bullet$ Feltre, A., et al., 2016, MNRAS 456, 3354;
$\bullet$ Franchini et al. 2024, A\&A, 686, 288;
$\bullet$ Haiman, Z., et al., 2023, arXiv:2306.14990; 
$\bullet$ Hammer F., et al., 2021, Msngr 182, 33;
$\bullet$ Ivezi{\'c}, {\v{Z}}., et al., 2019, ApJ, 873, 111;
$\bullet$ Izquierdo-Villalba, D., et al., 2023, A\&A, 677, 123;
$\bullet$ Krolik, J. H., 2010, ApJ 709, 774; 
$\bullet$ Krause, M., et al., 2025, PASA, 1;
$\bullet$ Krauth, L.M., et al. 2025, MNRAS, 543, 2670;
$\bullet$ Mainieri, V., et al., 2024, arXive:2403.05398;
$\bullet$ Mannucci, F., et al., 2023, A\&A, 680, 53;
$\bullet$ Mangiagli, A., et al., 2020, PhRvD, 102, 084056;
$\bullet$ Mazzolari, G., et al., 2024, A\&A 691, 345; 
$\bullet$ Murphy, E. J., et al., 2018, arXiv:1810.07524;
$\bullet$ Perna M., et al., 2025,  
$\bullet$ Prandoni, I. \& Seymour, N., 2015, aska.conf, 67;
$\bullet$ Reynolds, C., et al., 2023,  arXiv:2311.00780;
$\bullet$ Thatte, N., et al., 2021, Msngr 182, 7;
$\bullet$ Xin, C., \& Haiman, Z., 2021 
$\bullet$ Xin, C., et al. 2025, arXiv:2506.10846;

\end{document}